\documentclass[preprint]{elsarticle}

\usepackage[T1]{fontenc}
\usepackage[colorlinks,citecolor=blue,bookmarks=false,linkcolor=blue,urlcolor=blue]{hyperref}
\usepackage[english]{babel}
\usepackage[latin1]{inputenc}
\usepackage[normalem]{ulem}
\usepackage{amsmath}
\usepackage{amssymb}
\usepackage{caption}
\usepackage{chemist}
\usepackage{comment}
\usepackage{enumerate}
\usepackage{float}
\usepackage{gensymb}
\usepackage{graphicx}
\usepackage{hhline}
\usepackage{mathrsfs}
\usepackage{multicol}
\usepackage{multirow}
\usepackage{natbib}
\usepackage{revsymb}
\usepackage{siunitx}
\usepackage{subcaption}
\usepackage{supertabular}
\usepackage{textcomp}
\usepackage{url}
\usepackage{xcolor}

\journal{Computer Physics Communications}

\begin{document}

\begin{frontmatter}

\title{Benchmarking Collective Effects  of Electron Interactions in a Wiggler with OPAL-FEL}
\author[PSI]{Arnau Alb\`a}
\author[ANL,UNIST]{Jimin Seok}
\author[PSI]{Andreas Adelmann \corref{cor}}
\ead{andreas.adelmann@psi.ch}
\author[ANL]{Scott Doran}
\author[ANL]{Gwanghui Ha}
\author[ANL]{Soonhong Lee}
\author[ANL]{Yinghu Piao}
\author[ANL]{John Power}
\author[ANL]{Maofei Qian}
\author[ANL]{Eric Wisniewski}
\author[ANL]{Joseph Xu}
\author[ANL]{Alexander Zholents}

\affiliation[PSI]{organization={Paul Scherrer Institut},
            city={Villigen},
            postcode={5232},
            country={Switzerland}}

\affiliation[ANL]{organization={Argonne National Laboratory},
            city={Lemont},
            postcode={60439}, 
            state={Illinois},
            country={USA}}
            
\affiliation[UNIST]{organization={Ulsan National Institute of Science and Technology},
            city={Ulsan},
            postcode={44919}, 
            state={Gyeongnam},
            country={South Korea}}
            
\cortext[cor]{Corresponding author.}

\begin{abstract}
    OPAL-FEL is a recently developed tool for the modeling of particle accelerators containing wigglers or undulators. It extends the well established 3D electrostatic particle-tracking code OPAL, by merging it with the finite-difference time-domain electromagnetic solver MITHRA. We present results of two benchmark cases where OPAL-FEL simulations are compared to experimental results. Both experiments concern electron beamlines where the longitudinal phase space is modulated with a short magnetic wiggler. Good agreement was found in both the space charge and radiation dominated regimes. 
\end{abstract}


\end{frontmatter}

\section{Introduction}
\label{sec:introduction}
Wiggler magnets, consisting of alternating polarity dipole magnets with a strong transverse magnetic field, are used in electron/positron storage rings as devices for producing intense synchrotron radiation. See \cite{Winick, Kulipanov, Clarke} and references therein for an in depth introduction to the wigglers and undulators and descriptions of their diverse applications.
Coherent synchrotron radiation (CSR) in wiggler magnets can cause a microwave-like beam instability in storage rings as suggested in \cite{StupakovWig} and evidenced in \cite{ByrdPhysRevLett}. The observations in \cite{AboBakrPhysRevLett.88.254801,ArpPhysRevSTAB.4.054401,CarrNature,Podobedov} may also be associated with a CSR-driven instability.   
These findings inspired the study of the longitudinal wakefield and
impedance effect in a wiggler due to CSR described in \cite{wu03rs}. With the advent of the free-electron lasers (FELs), wigglers began to be employed in setups aiming for enhanced self-amplified spontaneous emission \cite{Zholents_PhysRevSTAB.8.040701,Duris_PhysRevLett} and setups that explore the wiggler's CSR for electron microbunching and obtaining large and narrow peak current spikes \cite{MacArthur_PhysRevLett}. These current spikes cause strong longitudinal space charge forces as discussed in \cite{GELONI_WigglerWake} that are  particularly noticeable in the wiggler where the beam propagates with reduced average longitudinal velocity $\bar{v}_z=c\sqrt{1-1/\gamma^2_z}$, where $c$ is the speed of light and $\gamma_z=\gamma/\sqrt{1+K^2/2}$. Here $\gamma$ is the relativistic factor and $K=e B_w \lambda_w/2\pi mc^2$ is the wiggler parameter, where $e, m$  are the electron charge and mass, $\lambda_w$ is the wiggler's period, and $B_w$ is the wiggler's peak magnetic field. When $K^2>>1$, the wiggler strongly influences the longitudinal space charge force such that the frequency of plasma oscillations inside the electron bunch propagating through the wiggler become a factor of $\gamma/\gamma_z$ larger than the frequency of the plasma oscillations in a drift section \cite{GELONISpaceCharge} and, thus, the wiggler can be conveniently used to control the plasma oscillation frequency. This feature can be useful for microbunched electron cooling (see, \cite{RatnerMBEC,StupakovMBEC} and references therein), in which one can consider shortening the amplification cascades by replacing the drifts with wigglers. However, due to CSR, the evolution of the microbunched beam inside the wiggler is more complex than in the drift and must be analyzed considering simultaneously the radiation and space charge forces. To the best of our knowledge, this analysis can only be performed in the integral form  \cite{GELONISpaceCharge, StupakovCSRandSpaceCharge} or numerically using a code based on first-principle equations.

Many computer tools for modeling wigglers and undulators have been developed during the last 20 years due to the increasing number of FEL facilities (some examples \cite{biedron19993d, reiche2014update, freund2014minerva}). However, these codes often make assumptions to be computationally efficient at the cost of limiting their applicability to certain types of FELs (see \cite{FALLAHI2018192} for a detailed comparison of FEL codes). For the experiments considered in this paper, it was necessary to use the less common electromagnetic particle-in-cell (EM-PIC) codes, which solve the full inhomogeneous Maxwell equations, and also simulate particles. Examples of such codes are OSIRIS \cite{Osiris} and MITHRA \cite{FALLAHI2018192}. In this paper, we present and use OPAL-FEL \cite{OPALFEL}, a combination of the 3D electrostatic code OPAL \cite{OPAL} and MITHRA. This code accounts for 3D effects and solves first-principle equations, making it very attractive for modeling the microbunched electron beam in a wiggler, but, it must first be benchmarked to experimental results. In this paper, we discuss the result of benchmarking simulations performed using OPAL-FEL for two experiments, one at SLAC using a wiggler and a high energy beam at LCLS in the regime dominated by CSR, and one at ANL using a wiggler and a low energy beam at AWA in the regime dominated by space charge.

\section{OPAL-FEL}
Having computational tools capable of accurately modeling accelerators has become essential for the study of complex phenomena in beam physics. The open source framework
OPAL (Object Oriented Parallel Accelerator Library) \cite{OPAL} is a parallel, fully 3D code for the simulation of charged particle beams. In order to compute space charge forces within the bunch, OPAL solves the static Maxwell equations in a Lorentz-boosted frame co-propagating with the particle. In this model, the magnetic field vanishes and one is simply left with the Poisson equation which is efficiently solved on a mesh using Fast Fourier Transform (FFT) techniques. While this approach is sufficient for many problems in accelerator science, the coupling between electric and magnetic fields is lost due to the static approximation, thus OPAL cannot account for the propagation of electromagnetic waves.

Recently, we integrated MITHRA \cite{FALLAHI2018192} into OPAL so that OPAL is now able to simulate wigglers and undulators \cite{OPALFEL} in addition to its general capabilities for the simulation of charged particle beams. In this new flavor of OPAL, OPAL-FEL, the inhomogeneous Maxwell equations are solved by considering the potential wave equations using finite differences in time and space. One can then explicitly update the scalar and vector potentials on a mesh. At each time step the charge and current density from the particles are deposited on the discrete mesh-points, and the fields are interpolated from the mesh-points to the particle positions. The entire simulation is done in a Lorentz-boosted frame moving at the average speed of the bunch. In this way, the mesh can be smaller since it only needs to encapsulate the bunch rather than the whole undulator. 

With OPAL-t the accelerator is modeled from the gun up to the undulator, using the static solver and, when it reaches the fringe fields of an undulator or wiggler, the MITHRA solver in OPAL is used to solve the full Maxwell equations. When the bunch exits the wiggler OPAL automatically switches back to OPAL-t with its static solver and continues the simulation until the desired end-point.

\section{LCLS Benchmark} \label{sec:LCLSBenchmark}

Wigglers and undulators can be used in FELs, not only at the end of the linac to produce coherent radiation, but also in the upstream section of the linac to reshape the electron bunch phase space. In these cases, a wiggler and a laser pulse are overlapped to generate small energy modulations in the bunch, that can later be compressed into short bunches \cite{zholents_obtaining_2005, hemsing_beam_2014}.

\begin{figure}[H]
  \centering
  \includegraphics[width=.8\textwidth, trim=0 0 0 25, clip]{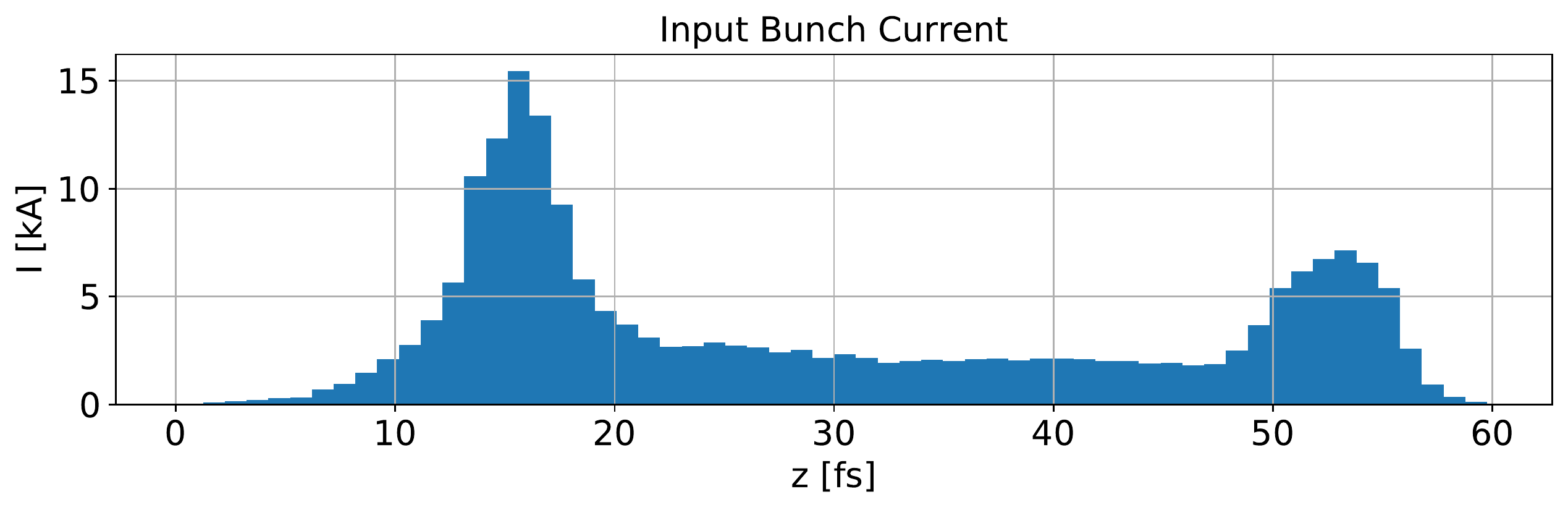}
  \caption{Initial bunch current for the LCLS simulation.}
  \label{fig_macArthur_current}
\end{figure}

In \cite{MacArthur_PhysRevLett}, MacArthur et al. investigate the possibility to generate a single cycle energy modulation in an electron bunch by means of a wiggler, but without the aid of an external laser. Using a line charge model they argue that a long bunch with a high-charge tail (figure \ref{fig_macArthur_current}), should emit coherent radiation strong enough to imprint an energy modulation in the beam core. In the paper they also validate their hypothesis through computer simulations with the electromagnetic code Osiris \cite{Osiris}, and with an experiment at LCLS.

The same experiment was simulated in OPAL-FEL to serve as a benchmark, as it represents an example where both space charge and radiation effects are important, and thus the full Maxwell equations need to be considered. The model included the 2.10 m wiggler, and an 80 cm drift before and after it, such that the initial and final bunch positions would be clear from the fringe fields. The input bunch had a charge of $Q=200$ \si{\pico\coulomb}, a transverse rms size $\sigma_{x,y}=74$ \si{\micro\metre}, an average energy of $E=3.95$ \si{\giga\electronvolt} with an energy spread of $\sigma_E=2$ \si{\mega\electronvolt}, and a current profile as shown in figure \ref{fig_macArthur_current}. The wiggler had a strength parameter $K=51.5$, and 6 periods of length $\lambda_w=35$ \si{\centi\metre}.

\begin{figure}[H]
  \centering
  \includegraphics[width=.8\textwidth]{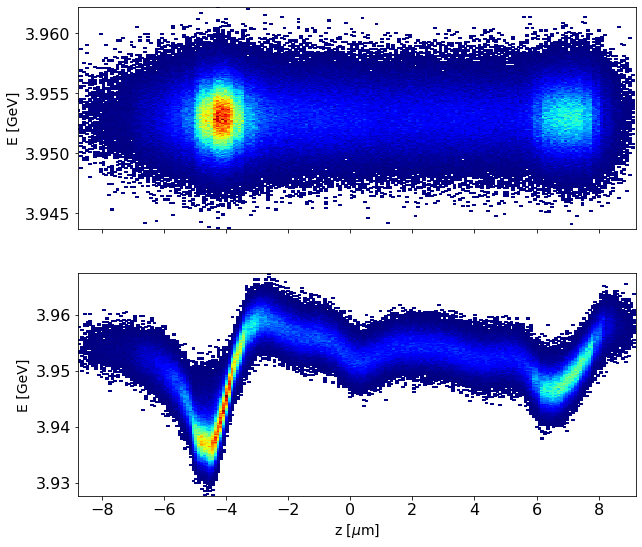}
  \caption{Longitudinal phase space of the electron bunch in the LCLS OPAL-FEL simulation (top) before the wiggler, at the starting point of the simulation, and (bottom) after the wiggler, where the phase space has been reshaped by wiggler-induced radiation.}
  \label{fig:macArthur_OPAL}
\end{figure}

\begin{figure}[H]
  \centering
  \includegraphics[width=.8\textwidth]{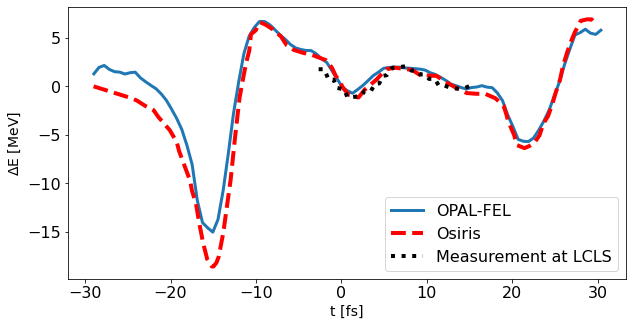}
  \caption{Comparison of the final slice-averaged energy between the OPAL-FEL simulation, and the OSIRIS simulation and measurement by MacArthur et al. \cite{MacArthur_PhysRevLett}.}
  \label{fig:OPAL_OSIRIS}
\end{figure}

The longitudinal phase space before and after the wiggler can be seen in figure \ref{fig:macArthur_OPAL}, where the wiggler-induced effects are clearly visible. The wake fields inside the wiggler affect mostly the high-charge spikes in the head and tail of the bunch. Less noticeably, the radiation also causes a chirp in the bunch center, as predicted by the theory. In figure \ref{fig:OPAL_OSIRIS} we see the slice-energy, where the single cycle energy modulation in the bunch center is evident. The slice-energy is also plotted for the OSIRIS simulation and the measurement at LCLS carried out by MacArthur et al, both showing good agreement with our OPAL-FEL simulations.

\section{AWA Experiment and Benchmark} \label{sec:AWABenchmark}

\subsection{Experimental Setup}

Among several available AWA beamlines \cite{conde2017research,ha2017limiting}, the straight section after the linac was used for this experiment (see figure \ref{fig:beamline_AWA}). The first 13 meters of the beamline form the photoinjector, consisting of a photocathode gun and six accelerating cavities. During this experiment, 4 accelerating cavities were used to generate 300 \si{\pico\coulomb} electron bunches with energy of 45.6 MeV. The straight section of the beamline that follows after the photoinjector is referred to as the "experimental beamline". It includes four matching quadrupoles (Q1-Q4, referred to as upstream quadrupoles), a wiggler, a longitudinal phase space (LPS) measurement section, and several YAG screens (diagnostic stations equipped with 50-mm diameter Cerium-doped Yttrium aluminium garnet (Ce:YAG or YAG for short) electron imaging screens to measure the transverse beam distribution) (fig. \ref{fig:beamline_AWA}). The upstream quadrupoles are used to match the beam from the photoinjector to produce the required transverse beam conditions at the wiggler center. The wiggler can be mechanically inserted in and out of the experimental beamline such that we were able to insert or remove the wiggler from the beam path (referred to as wiggler-in and wiggler-out settings). The LPS measurement section at the end of the beamline is used to measure the beam's temporal and energy spectrum perturbations after having passed through the wiggler. The LPS consists of four more quadrupole magnets (Q5-Q8, referred to as downstream quadrupoles) for transverse focusing at YAG5 and YAG6, a 100 \si{\micro\metre} horizontal slit for improved temporal resolution, a transverse deflecting cavity (TDC) and an energy spectrometer. The slit can be inserted or removed from the beamline when necessary.

The AWA photoinjector was used to generate a 300 \si{\pico\coulomb} electron bunch with an rms bunch length of around 0.1 mm. The photoinjector parameters to achieve this were determined with OPAL-t numerical simulations. The laser spot size (diameter) on the cathode was 12 mm and its pulse length (FWHM) was 300 fs. A single solenoid, located at the photocathode gun exit, was used to match bunches into the linac and on to the upstream quadrupoles. The solenoid was used to create a transverse waist at the entrance to the upstream quadrupole (i.e. Q1). The upstream quadrupoles were then easily used to match to the transverse beam conditions needed at the center of the wiggler. 

The measurement of the Twiss parameters at the wiggler center with wiggler-out was necessary for this experiment. However, the experimental beamline does not allow for the installation of a YAG screen at that point. For this reason we used a three screen method instead \cite{PhysRevAccelBeams.23.032804}, to measure the Twiss parameters. Once the upstream quadrupoles had been set to produce a specific beam condition, the three YAG screens (YAG1, YAG2, and YAG3), which are separated by drifts, were used to measure the rms beam sizes. We then analytically fit the beam sizes at the YAGs (wiggler-out), to obtain the Twiss parameters at YAG1 and at the wiggler center.

The main measurement necessary for this experiment was that of the longitudinal phase space. Since we know that the wiggler perturbs the beam's temporal and energy spectrums, we needed to compare the LPS of the wiggler-in to the LPS of the wiggler-out (and to our simulations). Downstream of the wiggler, the beam passes through the four downstream quadrupoles which are used to minimize transverse contributions on YAG5 and YAG6 \cite{PhysRevAccelBeams.21.062801}, where the longitudinal measurements are made. The 100 \si{\micro\metre} horizontal slit can be used in some cases to truncate the beam vertically to improve temporal resolution. The TDC deflects the beam in the vertical direction such that the temporal information gets projected onto the YAG in vertical direction. Energy spectrum information appears on the YAG in horizontal direction because the 20 degree energy spectrometer horizontally bends the beam depending on electron's energy. The LPS measurement section is able to measure the temporal distribution and the energy spectrum and the LPS, separately or simultaneously. In the case of temporal distribution only measurement (TDC-only), we focused the beam vertically at YAG5 and turned on the TDC (without the slit). Similarly, the energy spectrum (spectrometer-only) was measured with horizontal beam focusing at YAG6 with the energy spectrometer turned on (also without the slit). Lastly, the LPS measurement required transverse beam focusing at YAG5, the horizontal slit, and turning on both the TDC and the energy spectrometer. Note that drift length from the spectrometer to YAG5 and to YAG6 is the same.

The experiments and measurements were carried out for two specific settings of the upstream quadrupoles. They were set such that the transverse beam size would have a waist at the center of the wiggler. The two settings are from here on referred to as \textit{round beam} and \textit{elliptic beam}, and only differ in their transverse size, described in table \ref{tab:two_beams}.

\begin{table}[H]
    \centering
    \begin{tabular}{l c c}
         & $\sigma_x$ [mm] & $\sigma_y$ [mm]\\
        \hline
         Round Beam & 0.4 & 0.4\\
         Elliptic Beam & 2.5 & 0.4\\ 
    \end{tabular}
    \caption{Beam size at the center of the wiggler for the two beam settings.}
    \label{tab:two_beams}
\end{table}

The choice of the two transverse beam sizes was made in an attempt to perform an experiment that probed both above and below the wiggler's radiation diffraction limit $\sigma_{diff}$, given by

\begin{equation}
    \sigma_{diff} = \sqrt{\sigma_z\frac{\lambda_w}{2\pi}},
\end{equation}
where $\sigma_z$ is the longitudinal rms size of the bunch, and $\lambda_w$ the wiggler period. From \cite{GELONI_WigglerWake} we expect that with a transverse beam size $\sigma_r\gg\sigma_{diff}$ the wiggler-induced radiation effects become negligible, and only the space charge effects remain. In the current experiment, with an average bunch length of $\sigma_z = (250\pm40)$ \si{\micro\metre} and a wiggler period $\lambda_w=8.5$ \si{\centi\metre}, we find $\sigma_{diff} = (1.84\pm.15)$ \si{\milli\metre}. For this reason the horizontal beam size $\sigma_x$ was chosen to be in one case greater and in one case smaller than $\sigma_{diff}$. The vertical beam size $\sigma_y$ was unfortunately limited by the height of the vacuum chamber in the wiggler, and thus remained below the diffraction limit at all times.

\subsection{OPAL-FEL Benchmark}

\begin{figure}[H]
    \centering
    \includegraphics[width=0.98\textwidth]{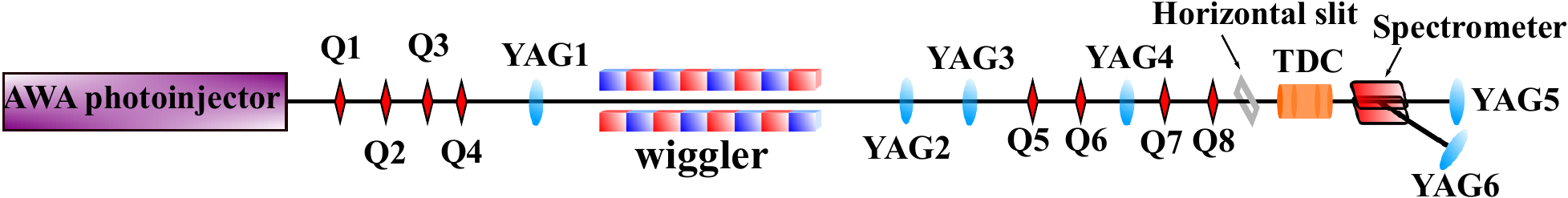}
    \caption{Experimental beamline at AWA. Note that the wiggler and slit can be inserted or removed from the beamline when required. The shown configuration is the wiggler-in case; in the wiggler-out case a simple drift replaces the wiggler.}
    \label{fig:beamline_AWA}
\end{figure}

In order to benchmark OPAL-FEL with the AWA wiggler experiment, the experimental beamline was computationally modeled and used for simulations that could be compared to the experiment. The simulated section goes from the first electron imaging screen (YAG1) up to the energy spectrometer 7 meters downstream. This section includes the wiggler, four YAGs for transverse beam measurements, the four downstream quadrupoles, and the LPS measurement section (figure \ref{fig:beamline_AWA}).

Simulating the electron bunch starting from YAG1, requires accurate knowledge of the bunch parameters in all 3 phase space planes at YAG1. The transverse beam parameters were known, since the transverse emittance and Twiss parameters at YAG1 were measured by the three-screen methods using YAG1-3. These transverse beam parameters can be seen in table \ref{tab:transverse_beam_params}.

\begin{table}[h]
    \centering
    \begin{tabular}{l c c c c c c}
         & $\beta_x$ [\si{\metre}] & $\alpha_x$ [ ] & $\epsilon_x$ [\si{\micro\metre}] & $\beta_y$ [\si{\metre}] & $\alpha_y$ [ ] & $\epsilon_y$ [\si{\micro\metre}]   \\
         \hline
         Round beam & 3.83 & -1.74 & 34.1 & 4.37 & -1.19 & 14.4 \\
         Elliptic beam & 20.0 & -0.23 & 31.7 & 2.84 & -0.93 & 17.1\\ 
    \end{tabular}
    \caption{Twiss parameters at YAG1 as measured during the experiment.}
    \label{tab:transverse_beam_params}
\end{table}

\begin{figure}[H]
    \centering
    \includegraphics[width=0.99\textwidth]{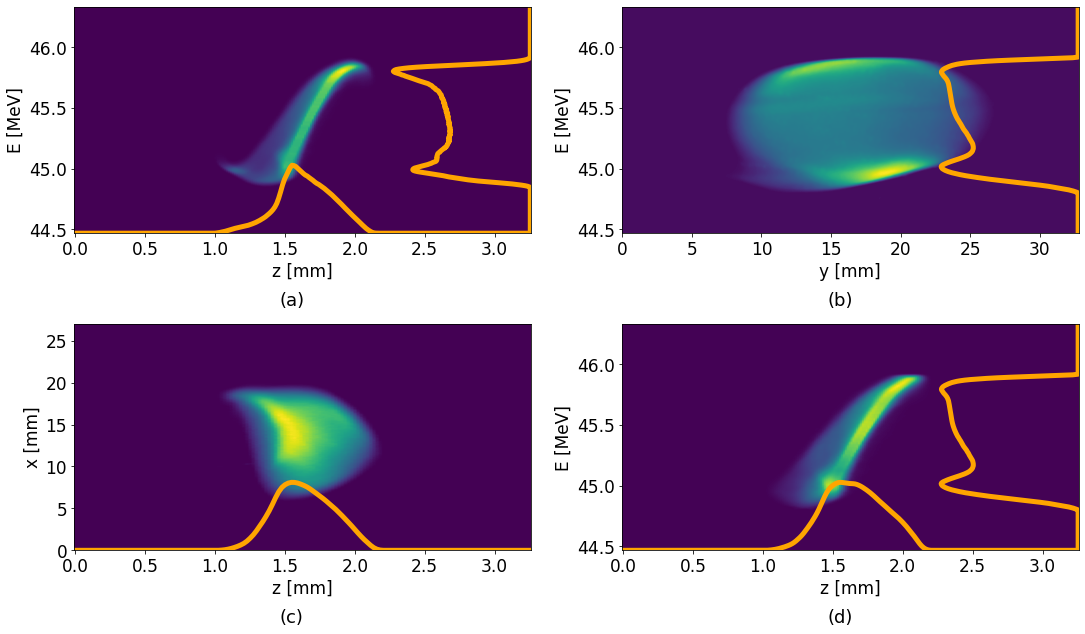}
    \caption{(a to c) Experimental YAG images taken with the wiggler out of the beamline, in the case of the round beam. All images show a bunch with charge $Q=(300\pm4)$ \si{\pico\coulomb}. (a) LPS observed on YAG6 with the TDC and spectrometer turned on. For this image a horizontal slit was placed in front of the TDC. (b) Energy distribution of the beam observed on YAG6, obtained with the spectrometer turned on. (c) Longitudinal particle distribution observed on YAG5 with the TDC turned on. (d) Numerical particle distribution generated by combining images a, b and c, which was used for the back-tracking simulations.}
    \label{fig:LPS_shots}
\end{figure}

The longitudinal beam parameters were numerically recreated at YAG1 by the following method.  The LPS was not directly known at YAG1, as it was only measured at the LPS measurement section at the end of the beamline. There, the spectrometer and TDC provided images of the longitudinal particle distribution and the energy spread, of which examples can be seen in figure \ref{fig:LPS_shots}. These images are available for both the round and elliptic beam experiments. Additionally, in each case, the experiment and images were repeated with the wiggler in the beamline (wiggler-in), and the wiggler removed from the beamline (wiggler-out). The images from the wiggler-in experiments were used later on for benchmarking the simulations, while the wiggler-out images were used to numerically recreate the LPS at YAG1.  In this way, the longitudinal parameters at YAG1 were combined with the Twiss parameters at YAG1 to obtain the initial bunch conditions for the simulations.

\begin{figure}[H]
    \centering
    \includegraphics[width=0.99\textwidth]{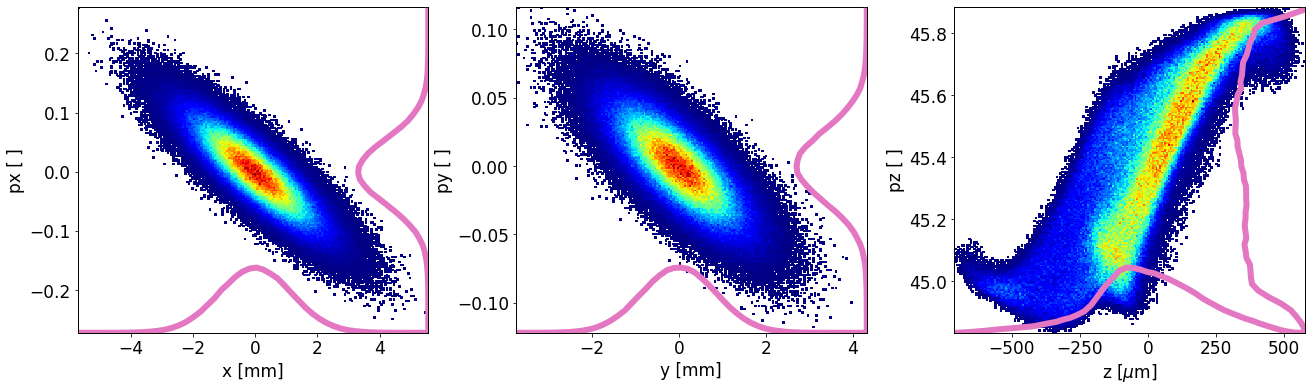}
    \caption{Initial particle distribution for the simulations of the round beam. This is the bunch's 6D phase space at YAG1.}
    \label{fig:distro_PYG3}
\end{figure}

In summary, the following steps were carried out to recover the full 6D phase space at YAG1:
\begin{enumerate}
    \item A 2D particle distribution was generated with the energy spread from the spectrometer images (figure \ref{fig:LPS_shots}b), and with the longitudinal distribution from the TDC images (figure \ref{fig:LPS_shots}c). The correlation of this 2D distribution was taken from the $z-E$ correlation seen in the LPS images (figure \ref{fig:LPS_shots}a), which were captured with both the spectrometer and TDC simultaneously turned on.  The resulting 2D distribution was the LPS at the spectrometer (figure \ref{fig:LPS_shots}d). All the images used in this step were taken with the wiggler-out. Note that the energy spectrum from the spectrometer shows a "bump" at $45.6$ MeV of energy (figure \ref{fig:LPS_shots}b). The source of this anomaly was discovered to be a low charge satellite bunch that was spuriously emitted from the cathode. This satellite bunch trailed the main bunch $4.5$ mm behind it, and had a charge of approximately $6$ pC. The effect of this low charge bunch is negligible with respect to the $300$ pC main bunch, and was thus not included in the simulations.
    \item The Twiss parameters at the spectrometer were computed by transferring the Twiss parameters from YAG1 (table \ref{tab:transverse_beam_params}), through the 7 meter section of the beamline to the spectrometer in the wiggler-out case, using linear transfer matrices.
    \item With the known Twiss parameters at the spectrometer, it was possible to sample the 4D transverse phase-space, and add it to the 2D longitudinal phase space generated in step 1, giving us the full 6D particle distribution. Note that with this method there were assumed to be no longitudinal-transverse correlations in the bunch.
    \item Finally, since OPAL's electrostatic solver assumes conservative forces, the simulation can be run in reverse to effectively track the bunch back in time. Using this approach, the bunch was simulated backwards from the spectrometer to YAG1, with the wiggler-out, thus obtaining the 6D particle distribution at YAG1 (figure \ref{fig:distro_PYG3}), the starting point of our simulations.
\end{enumerate}

With these four steps we were able to obtain the initial bunch conditions which were needed for our wiggler-in simulations. 

Note that simulation back-tracking only works in the wiggler-out case, because only the electrostatic solver can be run in reversed time. The MITHRA solver that OPAL-FEL uses to track the bunch through a wiggler cannot be run backwards in time.

The reader might remark that the experimental shots that combined the TDC and spectrometer simultaneously (figure \ref{fig:LPS_shots}a) provide a full picture of the LPS, and these LPS images could have been used to generate the 2D longitudinal particle distribution, instead of combining the TDC-only and spectrometer-only shots as we did. However, the shot from figure \ref{fig:LPS_shots}a was captured with a horizontal slit present in the beamline, placed in front of the TDC (figure \ref{fig:beamline_AWA}). This slit blocked part of the bunch, potentially distorting the histograms. As a consequence, we only took the $z-E$ correlation from these images, but not the energy and $z$ histograms. In the TDC-only and spectrometer-only images (figures \ref{fig:LPS_shots}b and c), the slit was not present, thus all electrons reached the YAG screens.

\begin{figure}[H]
    \centering
    \includegraphics[width=.99\textwidth]{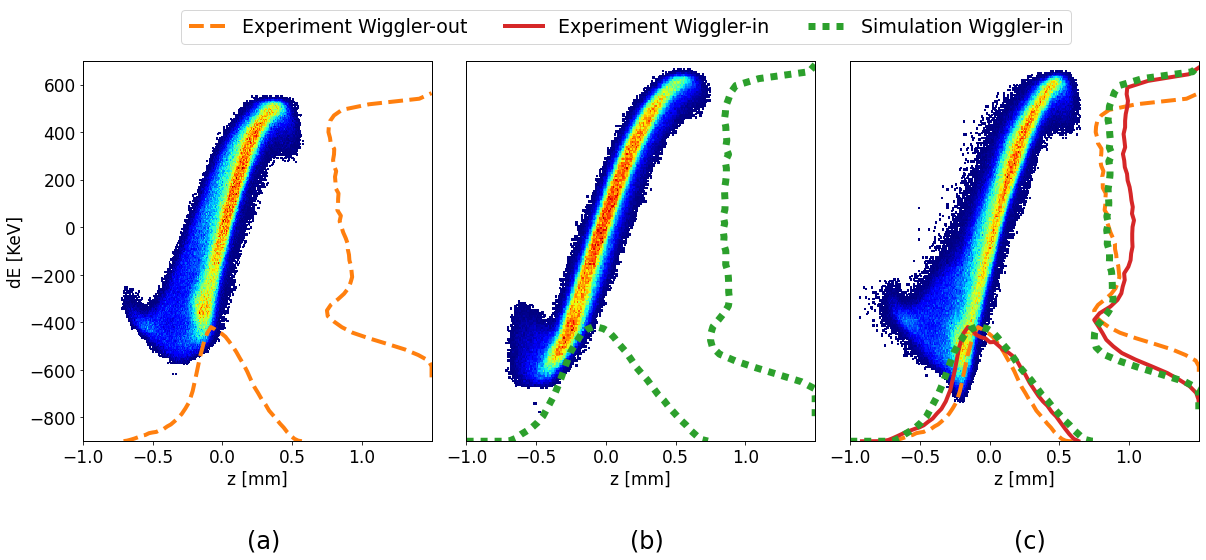}
    \caption{LPS of the round beam, at YAG6 in the (a) wiggler-out experiment, (b) the wiggler-in experiment, and (c) the simulation with wiggler-in.}
    \label{fig:AWA_comparison_round}
\end{figure}

\begin{figure}[H]
    \centering
    \includegraphics[width=.99\textwidth]{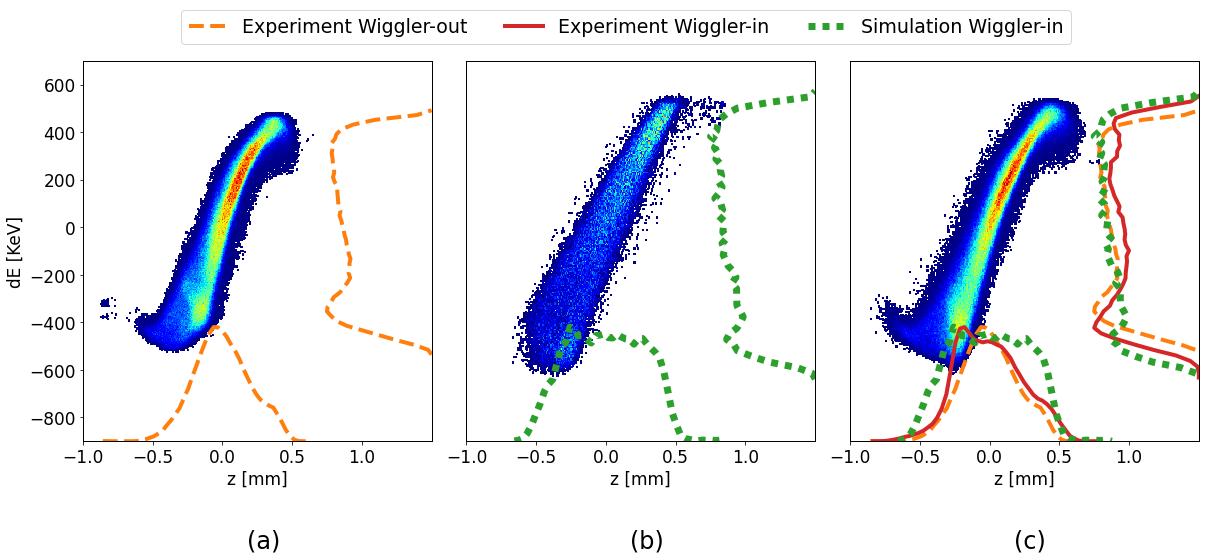}
    \caption{LPS of the elliptic beam, at YAG6 in the (a) wiggler-out experiment, (b) the wiggler-in experiment, and (c) the simulation with wiggler-in.}
    \label{fig:AWA_comparison_elliptic}
\end{figure}

With the known full particle distribution at YAG1, forward-time simulations were carried out with the bunch going through the wiggler and up to the spectrometer. The final LPS from the simulations can be compared to the experimental wiggler-in LPS shots (figures \ref{fig:AWA_comparison_round},\ref{fig:AWA_comparison_elliptic}). 

The most notable effect that the wiggler has on the bunch, as can be seen by comparing the wiggler-out and wiggler-in shots, is an increase in the total energy spread. This effect was to be expected since, as explained in the introduction \ref{sec:introduction}, the wiggler increases the plasma oscillation frequency of the bunch by enhancing space charge \cite{GELONISpaceCharge}. The same effect is also observed in simulations, as is apparent from the plots and table \ref{tab:FWHM_E}, where the full-width at half-maximum (FWHMs) of the energy spreads are reported.

\begin{table}[H]
    \centering
    \begin{tabular}{l c}
         & FWHM$_E$ [MeV] \\
         \hline
        Round beam, wiggler-out, exp &  $1.01\pm0.03$\\
        Round beam, wiggler-in, exp &  $1.24\pm0.03$\\
        Round beam, wiggler-in, sim &  $1.22\pm0.04$\\
        \hline
        Elliptic beam, wiggler-out, exp &  $0.91\pm0.03$\\
        Elliptic beam, wiggler-in, exp &  $1.08\pm0.03$\\
        Elliptic beam, wiggler-in, sim &  $1.03\pm0.04$\\
    \end{tabular}
    \caption{Values for the FWHM of the energy histograms observed on YAG6, from simulations and experiments.}
    \label{tab:FWHM_E}
\end{table}

Despite the good agreement between simulations and experiments, small discrepancies remain between the shape of the histograms (figure \ref{fig:AWA_comparison_round}c). These differences are suspected to be due to inaccuracies in the bunch's initial conditions for the simulations. The initial particle distributions were deduced from LPS measurements which had uncertainties, and gave no information about the transverse beam parameters, thus not allowing for an entirely realistic modeling of the bunch.

\section{Conclusion and Further Work
}

Good agreement with the experimental results has been demonstrated in the modeling of two experiments using OPAL-FEL. The considered experiments allowed us to test the code in different regimes of electron-wiggler interaction: a higher-energy case (c.f.\ section \ref{sec:LCLSBenchmark}) governed by the effects of emitted radiation, and a lower-energy case (c.f.\ section \ref{sec:AWABenchmark}) governed by space charge effects.
We have now all reasons to believe that the code can be reliably used for modeling of complex beamlines consisting of magnetic chicanes, wigglers, and undulators simultaneously accounting for the space charge and coherent synchrotron radiation effects.

With the OPAL-FEL code and thanks to the parallel nature of OPAL, full start-to-end modeling of complex beamlines is possible. In regions where radiation does not play a significant role, the 3D electrostatic solver is used, while in undulators and wigglers the full 3D electromagnetic solver will take radiation into account. The ability of OPAL-FEL to switch between the two solvers reduces the computational expense of simulations by using the faster electrostatic solver in the sections where radiation can be neglected. Adapting the numerical model to the physics needs has a great positive impact on the time to solution and contributes to an economical use of computing resources. 

The current plan is to apply OPAL-FEL to evaluate the gain in the recently proposed wiggler enhanced plasma cascade amplifier \cite{WEPA} for coherent electron cooling \cite{Litv_PhysRevLett.102.114801}. It is expected to be particularly fruitful since both the space charge and coherent synchrotron radiation effects are considered  to play equally important roles there.

\section{Statement of Contribution}
{\bf Arnau Alb\`a} integrated OPAL and MITHRA, carried out simulations, analyzed the experimental data and lead preparation of the paper for a publication; {\bf Jimin Seok} conducted the experiment, analyzed the data and contributed to paper writing; {\bf Andreas Adelmann} provided ideas for OPAL-FEL, provided the oversight for the data analysis, and contributed to paper writing;  {\bf Gwanghui Ha} conducted the experiment;  {\bf Soonhong Lee, Maofei Qian, and Joseph Xu} designed the wiggler and provided the oversight for a fabrication of the  wiggler's strong back; {\bf Yinghu Piao} carried out magnetic measurements and wiggler's tuning; {\bf Scott Doran} designed the wiggler's vacuum chamber and support, installed the wiggler, and  assembled the beamline; {\bf Eric Wisniewski} assembled the beamline;  {\bf John Power} provided the oversight for the experiment and data analysis; {\bf Alexander Zholents} conceived the experiment, led the activity and contributed to paper writing.

\section{Acknowledgement}
We are grateful to John TerHAAR, Joseph Gagliano III, and Eric McCarthy from the Advanced Photon Source of Argonne National Laboratory who assembled the wiggler that worked flawlessly in the experiment. This work was supported by FY 2018 Research and Development for Next Generation Nuclear Physics Accelerator Facilities DOE National Laboratory Announcement Number: LAB 18-184 Proposal ID 0000235339. 
\bibliographystyle{elsarticle-num}
\bibliography{wiggler}

\end{document}